\documentclass[a4paper,aps,prl,floatfix,twocolumn,showpacs,nofootinbib,superscriptaddress]{revtex4-1}

\usepackage[dvips]{epsfig}
\usepackage{latexsym,amsmath,verbatim}
\usepackage{color}
\usepackage{hyperref}
\newcommand{\av}[1]{\langle {#1} \rangle}

\begin{document}

\title{Modeling human dynamics of face-to-face interaction networks}

\author{Michele Starnini} 

\affiliation{Departament de F\'\i sica i Enginyeria Nuclear,
  Universitat Polit\`ecnica de Catalunya, Campus Nord B4, 08034
  Barcelona, Spain}

\author{Andrea Baronchelli} 

\affiliation{Laboratory for the Modeling of Biological and Socio-technical
  Systems, Northeastern University, Boston MA 02115, USA}

\author{Romualdo Pastor-Satorras} 
\affiliation{Departament de F\'\i sica i Enginyeria Nuclear,
  Universitat Polit\`ecnica de Catalunya, Campus Nord B4, 08034
  Barcelona, Spain}

\date{\today}

\begin{abstract}
  Face-to-face interaction networks describe social interactions in
  human gatherings, and are the substrate for processes such as
  epidemic spreading and gossip propagation.  The bursty nature of
  human behavior characterizes many aspects of empirical data, such as
  the distribution of conversation lengths, of conversations per
  person, or of inter-conversation times. Despite several recent
  attempts, a general theoretical understanding of the global picture
  emerging from data is still lacking.  Here we present a simple model
  that reproduces quantitatively most of the relevant features of
  empirical face-to-face interaction networks. The model describes
  agents that perform a random walk in a two dimensional space and
  are characterized by an attractiveness whose effect is to slow down
  the motion of people around them.  The proposed framework sheds
  light on the dynamics of human interactions and can improve the
  modeling of dynamical processes taking place on the ensuing
  dynamical social networks.
\end{abstract}

\maketitle

Uncovering the patterns of human mobility \cite{gonzalez08} and social
interactions \cite{jackson2010social} is pivotal to decipher the
dynamics and evolution of social networks \cite{Newman2010}, with wide
practical applications ranging from traffic forecasting to epidemic
containment.  Recent technological advances have made possible the
real-time tracking of social interactions in groups of individuals, at
several temporal and spatial scales. This effort has produced large
amounts of empirical data on human dynamics, concerning letter
exchanges \cite{Oliveira:2005fk}, email exchanges \cite{barabasi05},
mobile phone communications \cite{gonzalez08}, or spatial mobility
\cite{Brockmann:2006:Nature:16437114}, among others.

Specially noteworthy is the data on face-to-face human interactions
recorded by the SocioPatterns collaboration \cite{sociopatterns} in
closed gatherings of individuals such as schools, museums or
conferences. SocioPatterns deployments measure the proximity patterns
of individuals with a space-time resolution of $\sim 1$ meter and
$\sim 20$ seconds by using wearable active radio-frequency
identification (RFID) devices.  The data generated by the SocioPattern
infrastructure show that human activity follows a bursty dynamics,
characterized by heavy-tailed distributions for the duration of
contacts between individuals or groups of individuals and for the time
intervals between successive contacts
\cite{10.1371/journal.pone.0011596,PhysRevE.85.056115}. 

The bursty dynamics of human interactions has a deep impact on the
properties of the temporally evolving networks defined by the patterns
of pair-wise interactions \cite{Holme2012}, as well as on the behavior
of dynamical processes developing on top of those dynamical networks
\cite{PhysRevE.85.056115,Stehle:2011nx,PhysRevE.83.025102,Lee:2010fk,Parshani:2010,albert2011sync,citeulike:1835858}.
A better understanding of these issues calls for new models,
capable to reproduce the bursty character of social interactions and
trace back their ultimate origin, beyond considering their temporal
evolution \cite{citeulike:10827892}. Previous modeling efforts mostly
tried to connect the observed burstiness to some kind of cognitive
mechanisms ruling human mobility patterns, such as a reinforcement
dynamics \cite{citeulike:9301798}, cyclic closure
\cite{journals/corr/abs-1106-0288} or preferential return rules
\cite{citeulike:7974615}, or by focusing on the relation between
activity propensity and actual interactions \cite{citeulike:10827892}.

In this Letter, we present a simple model of mobile agents that
captures the most distinctive features of the empirical data on
face-to-face interactions recorded by the SocioPatterns collaboration.
Avoiding any \textit{a priory} hypothesis on human mobility and
dynamics, we assume that agents perform a random walk in space
\cite{hugues95:_random_walks_I} and that interactions among agents are
determined by spatial proximity \cite{baronchelli2012consensus}.  The
key ingredients of the model are the following: We consider that
individuals have different degrees of social appeal or
\textit{attractiveness}, due to their social status or the role they
play in social gatherings, as observed in many social
\cite{citeulike:7631686}, economic \cite{hierarchy} and natural
\cite{Sapolsky05} communities.  The effect of this social
heterogeneity is that interactions, as well as the random walk motion
of the agents, are biased by the attractiveness of the peers they met
over time. Additionally, we assume, according to experimental data,
that not all the agents are simultaneously present in system, but can
jump in and out of an active state in which they can move and
establish interactions. We will see that these simple assumptions
allow the model to reproduce many of the properties of
face-to-face interaction networks.

\begin{figure}[tb]
  \includegraphics*[width=0.95\linewidth]{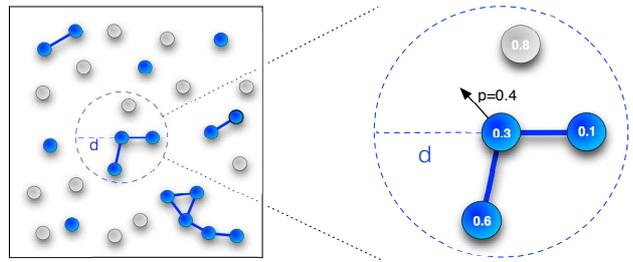}
  \caption{(color online) Left: Blue (dark) colored agents are active,
    grey (light) agents do not move nor interact. Interacting agents,
    within a distance $d$, are connected by a link.  Right: 
    Each individual is characterized by a
    number representing her attractiveness. The probability for the
    central individual to move is $p=1.0-0.6=0.4$, since the
    attractiveness of the inactive agent is not taken into account.  }
  \label{fig:rules}
\end{figure}

The model is defined as follows (see Fig.~\ref{fig:rules}): $N$  
agents are placed in a square box of linear size
$L$ with periodic boundary conditions, corresponding to a density $\rho= N/L^2$. 
Each individual $i$ is characterized by her attractiveness or social
appeal, $a_i$ which represents her power to raise interest in the
others. The attractiveness $a_i$ of the agents is a (quenched)
variable randomly chosen from a prefixed distribution $\eta(a)$, and
bounded in the interval $a_i \in [0,1)$.  Agents perform a random walk
biased by the attractiveness of neighboring individuals. Whenever an
agent intercepts, within a distance smaller than or equal to $d$,
another individual, they start to interact. The interaction lasts as
far as distance between them is smaller than $d$. Crucially, the more
attractive an agent $j$ is (the largest her attractiveness $a_j$), the
more interest she will raise in the other agent $i$, who will slow
down her random walk exploration accordingly. This fact is taken into
account by a walking probability $p_i(t)$ which takes the form:
\begin{equation}
\label{eq:1}
p_i(t) = 1- \max_{j \in \mathcal{N}_i(t) } \{ a_j \} ,
\end{equation} 
where $\mathcal{N}_i(t)$ is the set of neighbors of agent $i$ at time
$t$, i.e. the set of agents that, at time $t$, are at a distance
smaller than or equal to $d$ from agent $i$.  Hence, the biased random
walk performed by the agents is defined as follows: At each time step
$t$, each agent $i$ performs, with probability $p_i(t)$, a step of
length $v$ along a direction given by a randomly chosen angle $\xi \in
[0, 2\pi)$. With the complementary probability $1-p_i(t)$, the agent
does not move. Thus, according to Eq.~\eqref{eq:1}, if an agent $i$ is
interacting with other agents, she will keep her position in the
following time step with a probability proportional to the appeal of
his most interesting neighbor.

Furthermore, the empirical observations of SocioPatterns data show
that not all the agents involved in a social event are actually
present for its entire duration: Some agents leave the event before
the end, some join it later after the beginning, and some others leave
and come back several times.  Therefore we assume that agents can be
in an active or an inactive state. If an individual is active, she
moves in space and interacts with the other agents; otherwise she
simply rests without neither moving nor interacting.  At each time
step, one inactive agent $i$ can become active with a probability
$r_i$, while one active and isolated agent $j$ (not interacting with
other agents) can become inactive with probability $1-r_j$.  The
activation probability $r_i$ of the individual $i$ thus represents her
activeness in the social event, the largest the activity $r_i$, the
more likely agent $i$ will be involved in the event.  We choose the
activation probability $r_i$ of the agents randomly from an uniform
distribution $\zeta(r)$, bounded in $r_i \in [0,1]$, but we have
verified that the model behavior is independent of the activity
distribution functional form (even if we consider a constant activity
rate, $r_i = r $ for all agents, we obtain very similar results, see
Supplementary Material Figure 1).
 
Within this framework, each individual performs a discrete random walk
in a 2D space, interrupted by interactions of various duration with
peers. The movement of individuals is performed in parallel in order
to implement the time resolution ($20$ seconds) at which empirical
measurements are made \cite{10.1371/journal.pone.0011596}.  
The model is Markovian, since agents do
not have memory of the previous time steps.  The full dynamics of the
system is encoded in the collision probability $p_c = \rho \pi d^2$,
the activation probability distribution $\zeta(r)$, and the
attractiveness distribution $\eta(a)$.  The latter can hardly be
accessed empirically, and is likely to be in its turn the combination
of different elements, such as prestige, status, role, etc. Moreover,
in general attractiveness is a relational variable, the same
individual exerting different interest on different agents. Avoiding
any speculations on this point, we assume the simplest case of a
uniform distribution for the attractiveness
\cite{papadopoulos2012popularity}.  Remarkably, this simple assumption
leads to a rich phenomenology, in agreement with empirical
data.

\begin{table}[b]
  \begin{ruledtabular} 
    \begin{tabular}{|c||c|c||c|c||c|c|}
    Dataset & $N$ & $T$  & $\overline{p}$ &$\av{\Delta t}$ & $\av{k}$ & $\av{s}$ \\ \hline  
      hosp     &    84   & 20338 & 0.049  & 2.67 & 30 & 1145 \\
      ht         &    113 & 5093   & 0.060  & 2.13 & 39 & 366 \\ 
      school  &  126   & 5609   & 0.069  & 2.61 & 27 & 453 \\ 
      sfhh     &  416   & 3834   & 0.075  & 2.96 & 54 & 502\\ 
    \end{tabular}
  \end{ruledtabular} 
  \caption{Some properties of the SocioPatterns datasets under
    consideration: $N$, number of different individuals engaged in interactions;
    $T$, total duration of the contact sequence, in units of the
    elementary time interval $t_0 = 20$ seconds; 
    $\overline{p}$, average number of individuals interacting at each time step;
    $\av{\Delta t}$, average duration of a contact;
    $\av{k}$ and $\av{s}$: average degree and average strength of the
    projected network, aggregated over the whole sequence (see main text).} 
  \label{tab:summary}
\end{table}

In the following we will contrast results obtained by the numerical
simulation of the model against empirical results from SocioPatterns
deployments in several different social contexts: a Lyon hospital
(``hosp''), the Hypertext 2009 conference (``ht''), the Soci\`et\`e
Francaise d'Hygi\`ene Hospitali\`ere congress (``sfhh'') and a high
school (``school'').  A summary of the basic properties of the
datasets is provided in Table~\ref{tab:summary} (see
Refs.~\cite{10.1371/journal.pone.0011596,Isella:2011qo,PhysRevE.85.056115}
for further description and details).  The model has been simulated
adopting the parameters $v=d=1$, $L=100$ and $N=200$.  Different
values of the agent density $\rho$ are obtained by changing the box
size $L$. In the initial conditions, agents are placed at randomly
chosen positions, and are active with probability $1/2$. Numerical
results are averaged over $10^2$ independent runs, each one of
duration $T$ up to $T_\mathrm{max}=2 \times 10^4$ time steps.

\begin{figure}[tb]
  \includegraphics*[width=0.9\linewidth]{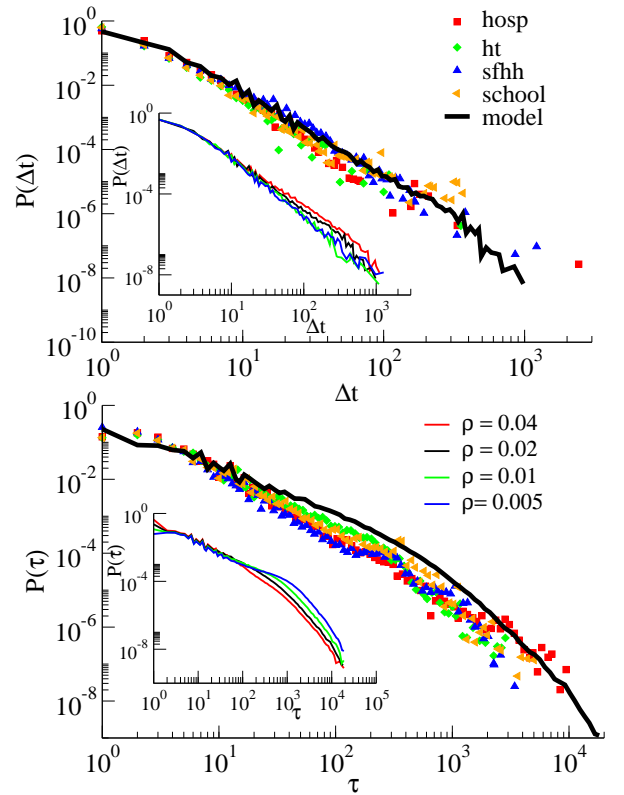}
  \caption{(color online) Distribution of the contact duration, $P(\Delta t)$, (up)
   and distribution of the time interval between consecutive contacts, $P(\tau)$, (down)
     for various datasets and for the attractiveness
     model.  Inset: Same distributions for
     the attractiveness model with different density $\rho$.  }
    \label{fig:duration} 
\end{figure}

The temporal pattern of the agents' contacts is probably the most
distinctive feature of face-to-face interaction networks
\cite{10.1371/journal.pone.0011596,PhysRevE.85.056115}. We therefore
start by considering the distribution of the duration $\Delta t$ of
the contacts between pairs of agents, $P(\Delta t)$, and the
distribution of gap times $\tau$ between two consecutive conversations
involving a common individual, $P(\tau)$. The bursty dynamics of human
interactions is revealed by the long-tailed form of these two
distributions, which can be described in terms of a power-law function
\cite{10.1371/journal.pone.0011596}. Figure~\ref{fig:duration} 
show the distribution of the contacts duration
$P(\Delta t)$ and gap times $P(\tau)$ for the various
sets of empirical data along with the same distributions obtained by
simulating the model described above with density $\rho=0.02$.  In the
case of the contact duration distribution, numerical and experimental
data match almost perfectly, see
Fig.~\ref{fig:duration}(up). Moreover, numerical results are robust
with respect to variations of the collision probability $p_c=\pi d^2
\rho$, as shown in the inset. It also worth highlighting
the crucial role played by the heterogeneity of attractiveness $a_i$.
In fact, assuming it constant, $a_i=a$ (and neglecting excluded volume
effects between agents) our model can be mapped into a simple first
passage time problem \cite{Redner01}, leading to a distribution
$P(\Delta t) \sim (\Delta t)^{-3/2}$ with an exponential cut-off
proportional to $d^2/(1-a)$. The (non-local) convolution of the
exponential tails induced by the heterogeneous distribution of
attractiveness leads in our model to a power law form, with no
apparent cut-off, and with an exponent larger than $3/2$, in agreement
with the result observed in the SocioPatterns data.
Regarding the distribution of gap times, $P(\tau)$, the model also
generates a long-tailed form, which is compatible, although in this case 
not exactly equal, to the empirical data, see
Fig.~\ref{fig:duration}(down).  The behavior of the distribution
$P(\tau)$ yielded by the model is substantially independent of
the agent density $\rho$ also in this case, as shown in the inset.

\begin{figure}[tb]
  \includegraphics*[width=0.9\linewidth]{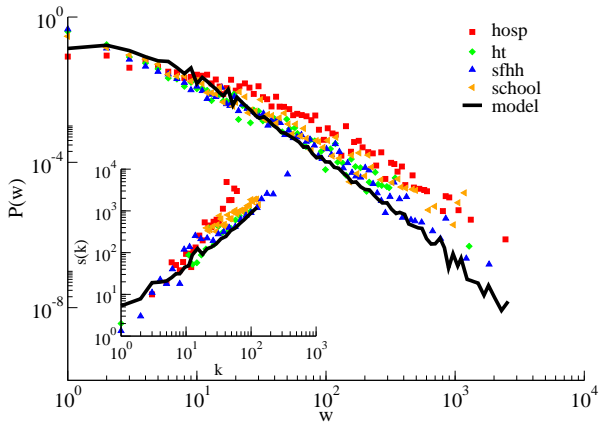}
  \caption{(color online) Weight distribution $P(w)$ (main) and 
  average strength of nodes of degree $k$, $s(k)$, as a function of $k$, (inset)
  for various empirical datasets and for the aggregate network obtained by
    simulating the attractiveness model.}
  \label{fig:w_distr} 
\end{figure}

Sociopatterns data can be naturally analyzed also in terms of
temporally evolving graphs \cite{Holme2012}, whose nodes are defined
by the agents, and whose links represent interactions between pairs of
agents. Instantaneous networks are thus formed by isolated nodes and
small groups of interacting individuals, not necessarily forming a
clique.  Integrating the information of these instantaneous graphs
over a time window $T$, which we choose here equal to the total
duration of the contact sequences defining each dataset
\cite{ribeiro2012quantifying}, produces an aggregated weighted network
\cite{Newman2010}, where the weight $w_{ij}$ between nodes $i$ and $j$
represents the total temporal duration of the contacts between agents
$i$ and $j$. The weight distribution $P(w)$ of the various datasets
are broad \cite{10.1371/journal.pone.0011596,PhysRevE.85.056115}, see
Fig.~\ref{fig:w_distr}(main), showing that the heterogeneity in the
duration of individual contacts persists even when contact durations
are accumulated over longer time intervals. Fig.~\ref{fig:w_distr}
shows that the outcome of the model is again in excellent agreement
with all empirical data, with the exception of the ``hosp''
database. The reason of the departure of this dataset with respect of
both other dataset and the model could be attributed to the duration
$T$ of the corresponding sequence of contacts (see
Table~\ref{tab:summary}), which is up to four times longer than the
other datasets.  In the limit of large $T$, sporadic interactions can
lead to a fully connected integrated network, very different from the
sparser networks obtained for smaller values of $T$. These effects
extend also to the pattern of weights, which have in the ``hosp''
database a much larger average value.

Face-to-face networks can be further characterized by looking at the
correlation between the number of different contacts and the temporal
duration of those contacts. These correlations be estimated by
measuring the strength $s_i$ of a node $i$, defined as $s_i = \sum_j
w_{ij}$ and representing the cumulated
time spent in interactions by individual $i$, 
as a function of its degree $k_i$, defined as the total different
agents with which agent $i$ has interacted.
Fig.~\ref{fig:w_distr}(inset) shows the growth of the average strength
of nodes of degree $k$, $s(k)$, as a function of $k$ in the empirical
datasets and in the aggregated network obtained with the
attractiveness model. As one can clearly see, all distributions (again
with the exception of the  ``hosp'' dataset) are well fitted
by a power law function $s(k) \sim k^{\alpha}$ with $\alpha > 1$, with
good agreement between real data and the model results. The observed
super-linear behavior implies that on average the nodes with high
degree are likely to spend more time in each interaction with respect
to the low-connected individuals \cite{10.1371/journal.pone.0011596}.

\begin{figure}[tb]
   \includegraphics*[width=0.9\linewidth]{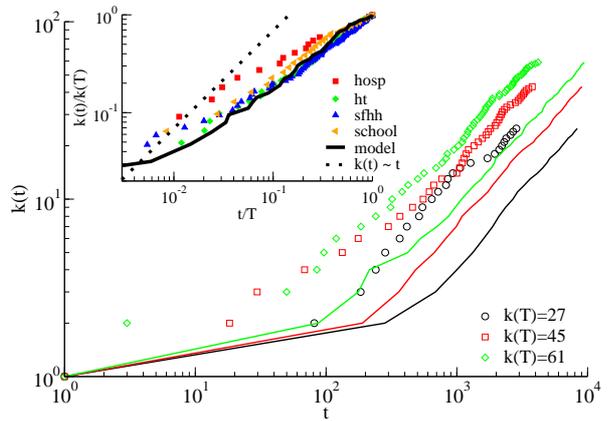}
   \caption{(color online) Main: Aggregated degree $k(t)$ versus time
     for various individuals with different final degree $k(T)$, for
     the ``ht'' dataset (symbols) and for the network
     obtained by simulating the attractiveness model (line). Inset:
     Rescaled aggregated degree $k(t) / k(T)$ as a function of time
     $t/T$ for various empirical datasets and for the attractiveness
     model. }\label{fig:k_t}
 \end{figure}

A final important feature of face-to-face interactions, also revealed
in different context involving human mobility
\cite{citeulike:7974615}, is that the tendency of an agent to interact
with new peers decreases in time. This fact translates into a
sub-linear temporal growth of the number of different contacts of a
single individuals (i.e. the aggregated degree $k_i(t)$), $k(t) \sim
t^{\mu}$, with $\mu<1$. Fig.~\ref{fig:k_t} shows the evolution of
$k(t)$ versus time for several agent with final aggregated degree
$k(T)$, both belonging to a single dataset (main) and for the
different datasets (inset). The sub-linear behavior of $k(t)$ is
clear, with $\mu = 0.6 \pm 0.15$ depending on the dataset. Moreover,
the shapes of the $k(t)$ functions can be collapsed in a single curve
by appropriately rescaling the data as $k(t)/k(T)$ as a function of
$t/T$, Fig.~\ref{fig:k_t}(inset). Fig.~\ref{fig:k_t} shows that,
remarkably, the attractiveness model is also capable to reproduce the
behavior of $k(t)$, up to the rescaling with total $T$ time, again
with the exception of the ``hosp'' dataset.


In summary, in this paper we have introduced a simple model of mobile
agents that naturally reproduces the social context described by the
Sociopatterns deployments, where several individuals move in a closed
environment
and interact between them
when situated within a small distance (the exchange range of RFID
devices). The main ingredients of the model are: (i) Agents perform a
biased random walk in two-dimensional space; (ii) their interactions
are ruled by an heterogeneous attractiveness parameter,
Eq.~\eqref{eq:1}; and (iii) not all agents are simultaneously active
in the system. Without any data-driven mechanism, the model 
is able to quantitatively capture most
of the properties of the pattern of interactions between agents, both
at level of the broad distributions of contact and inter-contact
times, and at the level of the ensuing temporal network.  Importantly,
results are robust with respect to variations of the model parameters,
i.e., the collision probability $p_c$ and the activity distribution
functional form, $\zeta(r)$. We have additionally checked that results
do not depend qualitatively on the nature of the motion rule, given by
Eq. \eqref{eq:1}. Indeed, other rules for the walking probability,
such as considering the average of the attractives of the neighbors,
i.e. $p_i(t) = 1- \sum_{j \in \mathcal{N}_i(t) } a_j / k_i(t)$, lead
substantially to the same behavior produced by Eq.~\eqref{eq:1} (see
Supplementary Material Figure 2). Overall, the proposed framework
represents an important step forward in the understanding of
face-to-face dynamical networks. Confronted with other modeling
efforts of SocioPatterns data \cite{citeulike:9301798},  
our model is not based on any cognitive
assumption (reinforcement dynamics in Ref.~\cite{citeulike:9301798}) 
and furthermore it leads to a good
agreement with experimental data without any fine tuning of internal
parameters.  It thus opens new interesting directions for future work,
including the study of dynamical processes taking place on
face-to-face networks and possible extensions of the model to more
general settings.

\begin{acknowledgements}
  We acknowledge financial support from the Spanish MEC, under project
  FIS2010-21781-C02-01, and the Junta de Andaluc\'{i}a, under project
  No. P09-FQM4682. R.P.-S. acknowledges additional support through
  ICREA Academia, funded by the Generalitat de Catalunya. We thank the
  SocioPatterns collaboration for providing privileged access to
 datasets ``hosp'', ``school'', and ``sfhh''. Dataset ``ht''
  is publicly available at~\cite{sociopatterns}.
  We thank A. Barrat for helpful comments and
  discussions.
\end{acknowledgements}


%

\end{document}